# Leveraging Machine Learning and Large Language Models for Automated Image Clustering and Description in Legal Discovery

Qiang Mao
Legal Technology & Data Analytics
Ankura Consulting Group, LLC
Washington, D.C. USA
qiang.mao@ankura.com

Fusheng Wei
Legal Technology & Data Analytics
Ankura Consulting Group, LLC
Washington, D.C. USA
fusheng.wei@ankura.com

Robert Neary
Legal Technology & Data Analytics
Ankura Consulting Group, LLC
Washington, D.C. USA
robert.neary@ankura.com

Charles Wang
Legal Technology & Data Analytics
Ankura Consulting Group, LLC
Washington, D.C. USA
charles.wang@ankura.com

Han Qin
Legal Technology & Data Analytics
Ankura Consulting Group, LLC
Washington, D.C. USA
han.qin@ankura.com

Jianping Zhang
Legal Technology & Data Analytics
Ankura Consulting Group, LLC
Washington, D.C. USA
jianping.zhang@ankura.com

Nathaniel Huber-Fliflet
Legal Technology & Data Analytics
Ankura Consulting Group, LLC
London, UK
nathaniel.huber-fliflet@ankura.com

*Abstract* - The rapid increase in digital image creation and retention presents substantial challenges during legal discovery, digital archive, and content management. Corporations and legal teams must organize, analyze, and extract meaningful insights from large image collections under strict time pressures, making manual review impractical and costly. These demands have intensified interest in automated methods that can efficiently organize and describe large-scale image datasets. This paper presents a systematic investigation of automated cluster description generation through the integration of image clustering, image captioning, and large language models (LLMs).

We apply K-means clustering to group images into 20 visually coherent clusters and generate base captions using the Azure AI Vision API. We then evaluate three critical dimensions of the cluster description process: (1) image sampling strategies, comparing random, centroid-based, stratified, hybrid, and density-based sampling against using all cluster images; (2) prompting techniques, contrasting standard prompting with chain-of-thought prompting; and (3) description generation methods, comparing LLM-based generation with traditional TF-IDF and template-based approaches.

We assess description quality using semantic similarity and coverage metrics. Results show that strategic sampling with 20 images per cluster performs comparably to exhaustive inclusion while significantly reducing computational cost, with only stratified sampling showing modest degradation. LLM-based methods consistently outperform TF-IDF baselines, and standard prompts outperform chain-of-thought prompts for this task. These findings provide practical guidance for deploying scalable, accurate cluster description systems that support high-volume workflows in legal discovery and other domains requiring automated organization of large image collections.

I. INTRODUCTION

The proliferation of digital imagery has created significant challenges in organizing, analyzing, and extracting meaningful insights from massive image collections. In legal discovery, practitioners routinely encounter datasets containing hundreds of thousands to millions of images that they must review, categorize, and understand under strict time and budget constraints [1]. Digital archives, e-commerce platforms, and research repositories face similar demands when making large visual collections accessible and interpretable. Manual review is prohibitively expensive, time-consuming, and prone to inconsistency and reviewer fatigue [2].

Recent advances in computer vision and natural language processing offer promising avenues for automated image analysis. Deep learning models now achieve strong performance in image classification, object detection, and visual feature extraction [3–4]. In parallel, large language models (LLMs) have demonstrated strong capabilities in understanding and generating human-like text, and emerging vision-language models increasingly bridge visual and textual modalities [5]. These developments point toward automated systems that can organize large image collections and generate meaningful textual descriptions of those collections.

Image clustering—the task of grouping visually similar images without predefined labels—has a long history in computer vision research [6]. Vision-language models capable of producing detailed image captions further enhance semantic understanding by converting visual content into textual descriptions [7]. This conversion enables the application of natural language processing techniques to image analysis. However, the specific task of generating human-interpretable descriptions for image clusters remains underexplored, despite its practical importance for domains that manage large image datasets.

This paper presents a systematic empirical investigation through three interconnected research questions:
(1) Sampling Strategy – *How do different image sampling strategies affect cluster description quality, and can strategic sampling match or exceed the performance of using all available images?*
(2) Prompt Engineering – *How do different prompting strategies influence the quality of generated cluster descriptions?*
(3) Method Comparison – *How do LLM-based description generation methods compare to traditional TF-IDF-based approaches in terms of description quality?*

## II. METHODOLOGY

Our methodology comprises three stages: (1) image clustering and captioning, (2) representative image sampling, and (3) cluster description generation. We first cluster all images using K-means and then generate captions for every image using a vision-language model. Next, we apply a range of sampling strategies to select representative images from each cluster. Finally, we generate cluster-level descriptions using both large language models and traditional TF-IDF–based methods to evaluate how sampling and prompting choices influence description quality.

### A. Image Clustering and Captioning

We first process all images through a pre-trained VGG16 convolutional neural network to extract high-dimensional feature vectors. Each image is encoded as a fixed-length vector that captures its key visual characteristics. We then apply K-means to partition the dataset into 20 clusters, using Euclidean distance to group images by visual similarity. The algorithm iteratively assigns images to clusters and updates centroid positions to minimize the within-cluster sum of squared distances.

To generate image captions, we use Azure AI Foundry's Image Analysis 4.0 Toolkit, powered by the Florence vision-language foundation model. For each image, the model produces up to ten regional captions describing distinct visual elements. We tokenize and clean these captions by removing stopwords before using them in downstream sampling and description generation tasks.

### B. Sampling Strategies

We compare six sampling approaches: random selection, centroid-based sampling (selecting images closest to the cluster center), stratified distance sampling (ensuring coverage across the distance distribution), hybrid sampling (combining multiple strategies), density-based sampling (focusing on high-density regions), and exhaustive inclusion of all images.

1. Random Sampling

Random sampling serves as our baseline approach, selecting $n$ images uniformly at random from each cluster without considering spatial distribution or visual similarity. We use random seed for reproducibility and select $n = 20$ images per cluster.

2. Centroid-Based Sampling

Centroid-based sampling prioritizes images most representative of the cluster by selecting those with the smallest Euclidean distance to the cluster centroid. This approach focuses on prototypical cluster members. Images are sorted by distance to centroid and 20 images per cluster were selected.

3. Stratified Distance Sampling

Stratified sampling divides the cluster into multiple strata based on distance quantiles from the centroid, then samples proportionally from each stratum to ensure comprehensive coverage across the entire distance spectrum. Distance quantiles are computed with images sampled proportionally in each stratum with 4 images per strata selected.

4. Hybrid Sampling

Hybrid sampling combines multiple strategies to balance representativeness and diversity. Our implementation allocates samples across three categories:

- Core samples (60%): Images closest to centroid

- Boundary samples (20%): Images farthest from centroid

- Random samples (20%): Randomly selected from mid-range distances

5. Density-Based Sampling

Density-based sampling uses kernel density estimation (KDE) to identify high-density regions in the distance distribution, then samples with probability proportional to local density. We apply Gaussian KDE to compute density $\rho_i$ for each image $i$ based on its distance from the cluster centroid, then sample with probability $p_i = \rho_i / \Sigma_j \rho_j$.

### C. Prompt Engineering

The quality of LLM-generated descriptions depends critically on how information is presented to the model. We evaluate two distinct prompting approaches: standard direct prompts and chain-of-thought (CoT) prompts that encourage step-by-step reasoning.

Standard prompt prioritizes efficiency and directness, allowing the model to leverage its pre-trained knowledge for pattern recognition without explicit reasoning steps. CoT prompting has shown improvements in complex reasoning tasks by making the model's reasoning process explicit [8]. Details of implementation for these two methods can be found in the appendix.

### D. Description Generation Methods

Beyond prompting strategies, we compare fundamentally different approaches to generating cluster descriptions: LLM-based generation versus traditional information retrieval methods.

1. LLM-Based Generation

Our primary approach leverages OpenAI's GPT-4o-mini model to generate natural language cluster descriptions. The model receives sampled image captions and produces coherent, human-readable descriptions that synthesize information across multiple images.

2. TF-IDF with Template-Based Generation

As a baseline comparison, we implement a traditional approach combining Term Frequency-Inverse Document Frequency (TF-IDF) keyword extraction with rule-based template generation. TF-IDF measures word importance by balancing term frequency within a cluster against its frequency across all clusters. We preprocess captions by tokenizing, lowercasing, removing stop words, and applying lemmatization to reduce words to their root forms. Each cluster is treated as a single document by concatenating all sampled captions, and TF-IDF scores are computed for all terms including both unigrams and bigrams. The top $k = 7$ keywords with highest scores are extracted and filtered by part-of-speech tags to choose only nouns. Cluster descriptions are then generated using rule-based templates that combine these selected nouns.

## III. EXPERIMENTS AND EVALUATIONS

### A. Experimental Setup

Our experiments were conducted on a collection of 16,240 images from a construction project, organized into 20 clusters using K-means clustering. Each image was then captioned using OpenAI's vision model.

All experiments were implemented in Python using scikit-learn for TF-IDF vectorization, NLTK for text preprocessing and POS tagging, and the OpenAI API for both image captioning and LLM-based description generation. For consistency across experiments, we used a fixed random seed for all sampling operations to ensure reproducibility. Cluster descriptions were generated using GPT-4o-mini with temperature set to 0.1 to promote consistent, factual outputs. Embedding computations for evaluation metrics utilized OpenAI's text-embedding-3-small model.

### B. Evaluation Framework

To evaluate the quality of automatically generated cluster descriptions, we adopt a semantic similarity-based evaluation framework that assesses how well generated descriptions align with the actual content of cluster members.

1. Semantic Similarity Calculation

Our evaluation relies on embedding-based semantic similarity to measure the alignment between cluster descriptions and individual image captions. The process consists of three steps:

(1) Embedding Generation
- Generate embeddings for all image captions using OpenAI's text-embedding-3-small model
- Generate embeddings for each cluster description
- All embeddings are L2-normalized unit vectors in a high-dimensional space

(2) Cosine similarity is used to calculate similarity between image caption embedding and cluster description embedding

(3) Aggregation Similarity scores are computed for all images within each cluster and aggregated to produce cluster-level and overall metrics.

This approach captures semantic alignment rather than lexical overlap, enabling evaluation of paraphrases, synonyms, and conceptually similar descriptions that may use different vocabulary.

2. Evaluation Metrics

We employ two primary metrics to assess description quality: Mean similarity Score and Coverage at 50% Threshold (Coverage@50).

Higher mean similarity indicates better alignment between descriptions and cluster contents. Values closer to 1.0 suggest the description accurately captures the common characteristics across cluster members, while lower values indicate weaker correspondence.

Coverage at 50% Threshold measures the percentage of images within a cluster whose captions have similarity ≥ 0.50 with the cluster description. It indicates what proportion of cluster members are reasonably well-represented by the description. Higher coverage indicates more comprehensive descriptions that apply broadly across the cluster.

These two metrics provide complementary perspectives: mean similarity assesses the strength of alignment, while coverage assesses the breadth of applicability.

### C. Results

1. Sampling Strategy Comparison

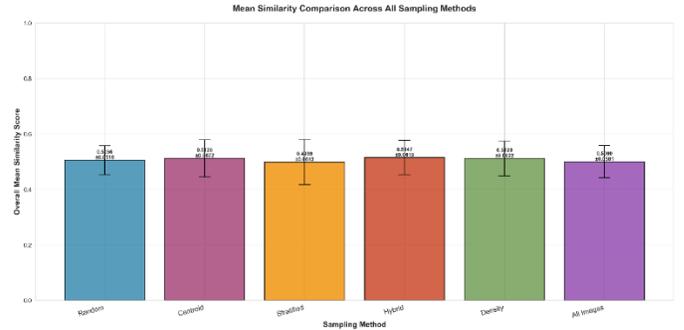

*Figure 1 Similarity Comparison Across All Sampling Methods*

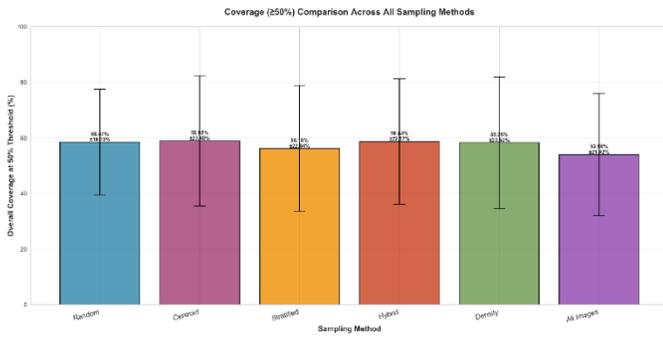

*Figure 2 Coverage@50 Comparison Across All Sampling Methods*

Figure 1 displays the mean similarity comparison, revealing that five of the six sampling methods showing marginal improvements over exhaustive inclusion. Stratified sampling is the only method performing slightly worse, though the difference is minimal. These results demonstrate that strategic selection of 20 images can match or exceed the performance of processing all available images.

Figure 2 presents the coverage@50 metrics, which measure the percentage of cluster images with similarity scores ≥ 0.50. All sampling methods achieve substantially higher coverage than All Images, showing approximately 2-5 percentage improvements. This pattern suggests that carefully selected subsets can actually provide better cluster representation than including all images, possibly because focused sampling reduces noise and redundancy that may dilute the description's relevance.

2. Prompt Engineering Comparison

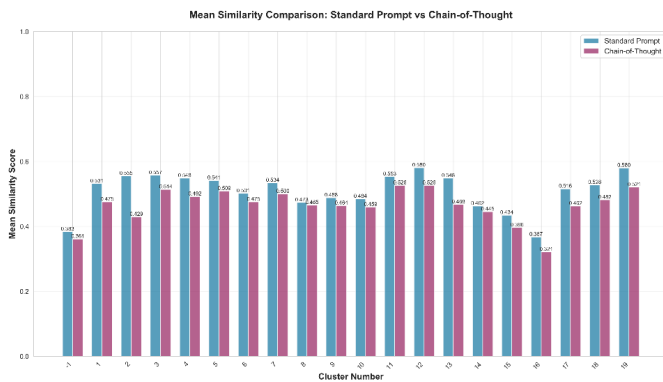

*Figure 3 Similarity Comparison: Standard Prompt vs. Chain-of-Thought*

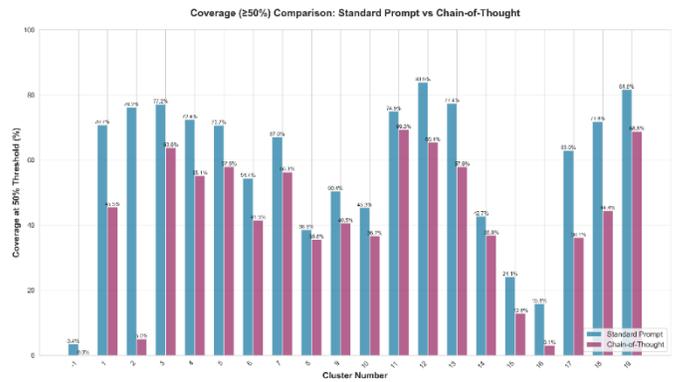

*Figure 4 Coverage@50 Comparison: Standard Prompt vs. Chain-of-Thought*

Figure 3 presents mean similarity scores across all 20 clusters for both prompting strategies using random sampling. Standard prompts consistently outperform chain-of-thought prompts across every cluster, demonstrating robust superiority regardless of cluster characteristics. Overall, standard prompts achieve a mean similarity of 0.51 compared to 0.46 for chain-of-thought prompts.

Figure 4 shows coverage@50 metrics, where the advantages of standard prompts become even more striking. Standard prompts achieve 58.1% overall coverage compared to 41.6% for CoT prompts. The performance gap varies considerably across clusters: Cluster 2 exhibits the largest disparity (76.2% vs. 5.0%). The substantial coverage advantage indicates that standard prompt descriptions align better with a broader proportion of cluster members, suggesting more comprehensive and representative characterizations.

3. Description Generation Method Comparison

Figure 5 presents mean similarity scores across all 20 clusters comparing LLM-based generation with TF-IDF + template-based generation, both using random sampling. LLM-based methods demonstrate clear superiority, outperforming TF-IDF in 19 of 20 clusters. Overall, LLM-based generation achieves a mean similarity of 0.51 compared to 0.46 for TF-IDF. The consistency of this advantage across nearly all clusters indicates that LLM-based synthesis provides systematically better semantic alignment with cluster contents.

Figure 6 displays coverage@50 metrics, where the performance gap between methods becomes even more pronounced. LLM-based generation achieves 58% overall coverage compared to only 36% for TF-IDF methods. Consistent with the mean similarity results, LLMs outperform TF-IDF in 19 of 20 clusters. Again, Cluster 16, which primarily features close-up views of documents, papers, and printed materials, is the exception where TF-IDF achieves superior coverage.

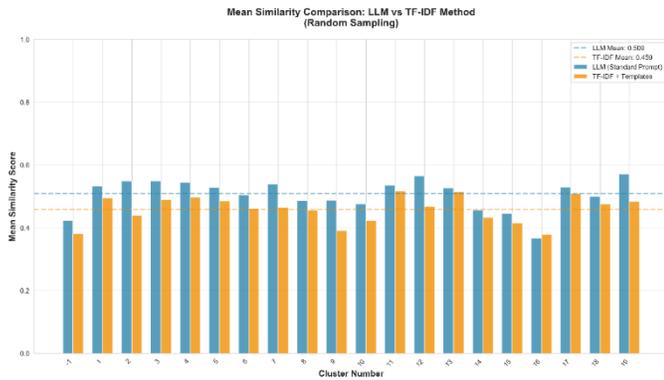

*Figure 5 Similarity Comparison: LLM vs. TF-IDF*

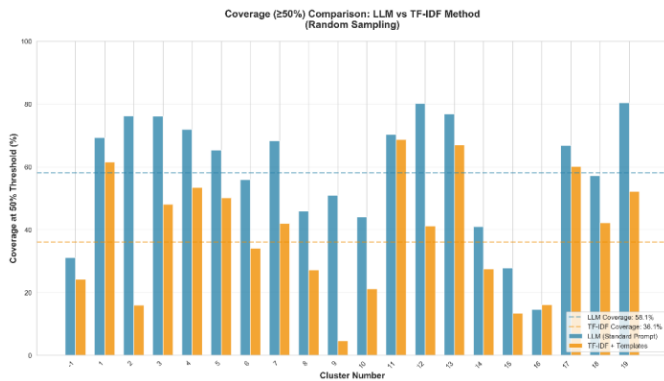

*Figure 6 Coverage@50 Comparison: LLM vs. TF-IDF*

## IV. CONCLUSIONS

This paper presents a systematic investigation of automated cluster description generation for large-scale image collections through the integration of K-means clustering, automated image captioning, and large language models. We examined three critical dimensions: sampling strategies for selecting representative images, prompt engineering approaches, and description generation methods. Our evaluation using semantic similarity metrics across image clusters provides practical guidance for implementing efficient and effective cluster description system.

Our experiments reveal three key insights. First, strategic sampling with only 20 images per cluster matches or exceeds the performance of exhaustive image inclusion across nearly all methods tested. Random, hybrid, density-based, and centroid-based sampling all achieve comparable or slightly superior results to using all images, with only stratified sampling showing worse performance. This finding demonstrates that thoughtful selection of a small subset can maintain description quality while reducing computational costs and API token usage. Second, standard direct prompts outperform chain-of-thought prompts in both semantic similarity and coverage metrics, demonstrating that simpler prompts offer superior efficiency in this case. Third, LLM-based description generation substantially outperforms traditional TF-IDF with template-based methods in both metrics, validating the value of modern language models for generating natural, accurate cluster descriptions.

This work addresses critical needs in legal discovery, digital asset management, content moderation, and scientific research where large image collections require efficient organization and interpretation. Our findings have immediate applicability across domains, and our results regarding sampling efficiency and prompt design contribute to broader understanding of effective LLM utilization beyond cluster description.

This research is limited to a single dataset containing approximately 16k images from a construction project. Validation across diverse datasets with varying sizes, domains, and visual content types is essential to strengthen the generalizability of our findings. A critical limitation is that cluster description quality is fundamentally constrained by the quality of clustering accuracy and caption precision. Future work should investigate robust quality assessment methods for clustering outputs and caption generation to ensure solid foundations before description generation. Additionally, prompt engineering remains an under-explored dimension in our work; systematic investigation of few-shot learning with carefully curated examples, alternative prompt formulations, and prompt optimization techniques could further enhance description quality.

## CONCLUSIONS

[1] Grossman, M. R., & Cormack, G. V. (2010). Technology-assisted review in e-discovery can be more effective and more efficient than exhaustive manual review. *Rich. JL & Tech.*, *17*, 1.
[2] Roitblat, H. L., Kershaw, A., & Oot, P. (2010). Document categorization in legal electronic discovery: computer classification vs. manual review. *Journal of the American Society for Information Science and Technology*, *61*(1), 70-80.
[3] He, K., Zhang, X., Ren, S., & Sun, J. (2016). Deep residual learning for image recognition. In *Proceedings of the IEEE conference on computer vision and pattern recognition* (pp. 770-778).
[4] Redmon, J., Divvala, S., Girshick, R., & Farhadi, A. (2016). You only look once: Unified, real-time object detection. In *Proceedings of the IEEE conference on computer vision and pattern recognition* (pp. 779-788).
[5] Radford, A., Kim, J. W., Hallacy, C., Ramesh, A., Goh, G., Agarwal, S., ... & Sutskever, I. (2021, July). Learning transferable visual models from natural language supervision. In *International conference on machine learning* (pp. 8748-8763). PmLR.
[6] Jain, A. K. (2010). Data clustering: 50 years beyond K-means. *Pattern recognition letters*, *31*(8), 651-666.
[7] Li, J., Li, D., Xiong, C., & Hoi, S. (2022, June). Blip: Bootstrapping language-image pre-training for unified vision-language understanding and generation. In *International conference on machine learning* (pp. 12888-12900). PMLR.
[8] Wei, J., Wang, X., Schuurmans, D., Bosma, M., Xia, F., Chi, E., ... & Zhou, D. (2022). Chain-of-thought prompting elicits reasoning in large language models. *Advances in neural information processing systems*, *35*, 24824-24837.

## APPENDIX A: PROMPTS

A. Standard Prompt

The standard prompt provides direct instructions for generating cluster descriptions without explicit reasoning steps:

1. System Prompt

*You are an expert analyst specializing in image analysis and pattern recognition.*
*Your task is to analyze collections of image captions and generate concise, accurate descriptions that capture the essential characteristics of image clusters.*

*Your descriptions should:*
*1. Identify the common theme or subject matter across the images*
*2. Be specific and factual, avoiding vague generalizations*
*3. Be concise (2-4 sentences maximum)*
*4. Highlight distinguishing features that differentiate this cluster from others*
*5. Focus on observable visual content and patterns*

2. User Prompt

*Analyze the following image captions from Cluster [ID]. These images were grouped together by a machine learning clustering algorithm based on visual similarity.*
*{IMAGE CAPTIONS}*
*Based on these captions, generate a comprehensive description of this image cluster that:*
*1. Identifies the primary subject matter or theme*
*2. Notes any consistent visual patterns or characteristics*
*3. Describes the context (e.g., type of scene, environment, objects, activities)*
*4. Mentions any notable commonalities across the images*

B. Chain-of-Thought (CoT) Prompt

The chain-of-thought prompt encourages the model to break down the analysis into explicit reasoning steps before generating the final description:

1. System Prompt

*You are an expert analyst specializing in image clustering and pattern recognition.*
*Use step-by-step reasoning to analyze image clusters and generate accurate descriptions.*

2. User Prompt

*Analyze Cluster [ID] using step-by-step reasoning. These images were grouped together by a clustering algorithm based on visual similarity.*
*{IMAGE CAPTIONS}*
*ANALYSIS STEPS:*
*1. First, identify common keywords and themes across all captions*
*2. Then, determine the primary content category or type of imagery*
*3. Next, note any patterns in visual characteristics, settings, or compositions*
*4. Consider what makes this cluster cohesive and distinct from other potential clusters*
*5. Finally, synthesize a concise description (2-4 sentences)*